# Can AI Be a Moral Victim? The Role of Moral Patiency and Ownership Perceptions in Ethical Judgments of Using AI-Generated Content

Hyesun Choung
Brian Lamb School of Communication
Purdue University
West Lafayette, Indiana, USA
choungh@purdue.edu

Soojong Kim
Department of Communication
University of California, Davis
Davis, California, USA
sjokim@ucdavis.edu

## Abstract

The growing use of generative AI raises ethical concerns about authorship attribution and plagiarism. This study examines how people judge the reuse of AI-generated content, focusing on moral patiency and ownership perceptions. In an experiment, participants evaluated two substantively similar manuscripts in which the original source was described as authored by a human, an AI system, or an AI agent with a human-like name. Results showed that copying AI-generated work was judged less unethical, less plagiaristic, and less guilt-inducing than copying human-authored work. Mediation analyses revealed that this leniency stemmed from lower perceptions of AI's capacity to suffer harm (moral patiency) and greater ownership attributed to the human writer reusing AI-generated content. Anthropomorphic cues shaped moral evaluations indirectly by reducing perceived ownership. These findings shed light on how people morally disengage when using AI-generated work and highlight differences in how ethical judgments are applied to human versus AI-created content.

## CCS Concepts

• **Human-centered computing** → Collaborative and social computing; Empirical studies in collaborative and social computing; • **Applied computing** → Law, social and behavioral sciences; Psychology.

## Keywords

Generative AI, Content ownership, Moral patiency, Ethics of AI

## 1 Introduction

Generative Artificial Intelligence (AI) tools are reshaping communication practices, from academic writing and professional communication to personal interactions and creative expression. These tools raise fundamental questions about authorship, ownership, and ethical responsibility in content creation. One emerging concern is "aigiarism," the act of plagiarizing AI-generated content [12, 35]. With AI increasingly serving as an invisible partner in content creation, AI-generated content blurs the boundaries of ownership and responsibility. Many generative systems operate within a mixed-initiative co-creative process, in which humans and AI systems iteratively contribute to the production of content. Users initiate this process through prompting and refinement, while AI systems respond with generated outputs that people evaluate, edit, and incorporate into their own work [17, 62]. This collaborative structure distributes perceived agency and control between human and computational actors, complicating questions of authorship and ethical responsibility.

While norms of attribution are generally clearer in human-authored work, AI-generated content introduces greater uncertainty about who owns the material and whether using it without credit is unethical. Users may feel justified in reusing AI outputs without credit, perceiving AI systems as lacking legitimate claims to intellectual property. Supporting this perception, a recent study found that people view plagiarism of AI-generated content as less unethical and more permissible than plagiarizing human-generated work because individuals do not naturally attribute ownership to AI [41].

The current legal framework for copyright, rooted in the protection of human authorship, struggles to offer definitive guidance on the ethical use of AI-generated content [13]. Unlike cases of plagiarism involving human-created work, where there is an identifiable victim, questions about who–or what–is harmed by the use of AI-generated content are ambiguous. While there may indeed be human creators whose original works have been incorporated into AI training data without consent, identifying such victims is challenging due to the large-scale and black-box nature of generative models, which obscures accountability [43, 48]. Additionally, the widespread adoption of generative AI tools like ChatGPT normalizes the use of AI for content creation, reducing individuals' sensitivity to potential ethical issues. Beyond questions of ownership, broader ethical debates surrounding generative AI involve concerns about transparency in disclosing AI involvement [50], the authenticity of AI-generated outputs [29], and the provenance of data used to train such systems [38]. These issues collectively shape the moral and regulatory landscape in which users form judgments about the ethical use of AI-generated content. In this context, the growing prevalence of AI-mediated communication practices [27] underscores the need to better understand how people morally evaluate and justify the use of AI-generated content.

These challenges posed by AI-generated content bring into focus differences in how humans perceive and interact with AI compared to other humans. Interactions between people are governed by established norms related to reciprocity, trust, and responsibility [11, 22, 42]. However, studies have shown that people perceive AI as lacking agency and emotional experience [39], both of which are critical for making ethical judgments [6, 40]. This perception diminishes users' sense of obligation to acknowledge AI for its output, making the moral concerns surrounding AI content less salient.

Such differences in moral evaluations can be explained by Mind Perception Theory [23], which posits that people attribute mental capacities to others along two dimensions: agency (the ability to act intentionally) and experience (the capacity to feel emotions and suffer harm). AI systems may sometimes be perceived as having agency due to their autonomy in decision-making, but they are often not seen as possessing experience or consciousness [57]. Because AI is not expected to experience emotions or engage in emotional reciprocity, it is not perceived as capable of suffering harm and, therefore, is not viewed as a moral victim. As a result, ethical considerations related to AI can differ substantially from those involving humans. For example, users may feel less guilt and perceive minimal harm when plagiarizing AI-generated content, precisely because AI lacks the moral patiency (the status of being a legitimate recipient of moral concern) necessary to be seen as a victim. Similarly, ownership perceptions are shaped by the belief that AI is merely a creative tool rather than a co-creator with legitimate authorship [60].

To gain insight into how individuals interact with generative AI tools for communication practices, this study examines the psychological underpinnings that shape moral judgments about AI-generated content. Specifically, we investigate how perceptions of moral patiency and ownership influence evaluations of plagiarism and associated feelings of guilt. Our focus is descriptive rather than normative: we examine how people's perceptions of AI's moral patiency, or lack thereof, shape their ethical judgments. Importantly, this study does not suggest that AI systems possess or should possess moral standing. We do not argue that AI can experience harm; rather, we investigate how the belief that AI *cannot* experience harm affects users' moral reasoning about authorship and plagiarism.

Our findings make three contributions: (1) we empirically demonstrate that plagiarism of AI-generated content is judged more leniently than plagiarism of human-authored work, (2) we extend theories of mind perception and ownership by showing how diminished perceptions of AI's moral patiency and heightened ownership attributions to human writers mediate these judgments, and (3) we highlight design implications by showing that anthropomorphic cues can alter users' ethical evaluations. Together, these contributions advance HCI research on AI-mediated communication, authorship, and responsible system design, offering conceptual and practical insights into how ethical norms evolve as generative AI becomes embedded in everyday writing and creative work.

## 2 Related work and Hypotheses
### 2.1 Generative AI and Moral Judgments: Psychological Underpinnings

Research has demonstrated that people can attribute mind perceptions to robots and other technologies, perceiving them as capable of agency and, in some cases, experience [58]. This insight aligns with the Computers as Social Actors (CASA) framework, which suggests that individuals unconsciously respond to computers and AI systems based on social psychological principles [10, 47, 49]. While these attributions influence human-machine interactions, they do not equate AI with human beings. People still recognize that non-human entities like machines, robots, and AI agents lack certain mental capacities, such as consciousness and intentional experience [24, 51]. As a result, researchers caution against applying human moral norms directly to AI systems, as they lack the essential mental states that anchor human moral judgments [21, 45].

Moral psychology seeks to understand how individuals make judgments about right and wrong, often emphasizing concepts such as responsibility, harm, and intent [52, 53]. These factors are crucial for ethical decision-making, particularly when evaluating actions like plagiarism. Moral judgment is often described as an intuitive process influenced by innate and culturally shaped moral foundations [25]. According to Moral Foundation Theory, common principles such as fairness, harm, and authority guide these judgments [26]. In cases of human plagiarism, ethical violations are typically judged based on harm caused to the author of the original content, including emotional distress, reputational damage, or lost opportunities. Copyright law is designed to prevent such harm by protecting intellectual property.

However, the application of these moral principles is far less straightforward in the context of AI-generated content. Mind Perception Theory [23] emphasizes how individuals attribute mental capacities to entities along two key dimensions: agency (the ability to act intentionally, plan, and make decisions) and experience (the capacity to feel emotions). These dimensions influence whether people view an entity as a moral agent (responsible for actions) or a moral patient (capable of being harmed, thus deserving moral concern and protection from harm). Drawing on this theory, Sullivan and Fosso Wamba [55] demonstrated that people apply different moral criteria when judging human versus AI actions. While human moral evaluations rely on both agency and experience, AI systems were often perceived as having insufficient experience, which leads to lower moral expectations regarding their treatment. Ladak et al. [39] found that while AIs were attributed low-to-moderate agency, they were generally seen as having very limited experience (often comparable to inanimate objects such as a rock). Because moral patiency was closely tied to perceived experience, even the most experiential AI (a robot dog) was rated far below the least experiential animal (an ant). Taken together, these findings suggest that people largely withhold moral concerns from AIs due to their perceived lack of inner experience, which diminishes beliefs that AI systems can be harmed and weakens ethical accountability in interactions with them.

For instance, Longoni et al. [41] found that participants were more willing to submit AI-generated essays under their own name than content plagiarized from a human source. Participants also attributed greater ownership to themselves when using AI-generated content, suggesting that AI is not viewed as having any legitimate claims to authorship. This perception reduces users' moral accountability, as they do not see their actions as harmful or irresponsible when using AI content without acknowledgment. Consequently, relying on users' intrinsic sense of moral wrongness to deter plagiarism in AI contexts is likely ineffective because unlike human creators, AI systems are not perceived as capable of emotional harm or ownership, which undermines the relevance of both the care/harm and fairness foundations.

These perceptual gaps pave the way for moral disengagement, a process in which individuals rationalize unethical behavior by minimizing or dismissing the potential harm caused [7, 19]. In the context of AI-generated content, users may justify actions such as failing to credit AI outputs, reasoning that no conscious entity is capable of being harmed. Over time, this rationalization may become normalized, especially as generative AI tools like ChatGPT become seamlessly integrated into communication and creative practices [54]. As a result, users may become less sensitive to ethical issues surrounding authorship and intellectual integrity, ultimately influencing norms and expectations around the use of AI-generated content.

## 2.2 Moral Patiency and Ownership Perceptions of AI-Generated Content

In moral judgments surrounding the use of AI-generated content, ownership perceptions and moral patiency can play key roles in shaping users' evaluations of right and wrong. The concept of moral patiency involves recognizing an entity as capable of experiencing harm or benefit [8, 9]. In human-to-human interactions, moral norms often revolve around preventing emotional, reputational, or material harm to others. However, AI systems, which have not yet reached a sufficient level of consciousness and sentience, do not meet the criteria for moral patiency [3]. Without these capacities, AI is not seen as a victim capable of being meaningfully affected by harm. Studies have shown that individuals experience reduced guilt and diminished moral concern when engaging in questionable actions involving AI, in part because they do not perceive AI as capable of being harmed [21, 36]. Based on these insights, we hypothesize:

- **H1:** Individuals will attribute lower moral patiency to an AI source than to a human source of the original content.

In addition to moral patiency, ownership perceptions can influence ethical evaluations of AI-generated content. Avey et al. [4] explained that individuals develop a sense of ownership over objects or outputs when they feel that they have control, contributed effort, or personal investment in the creation process. In AI content creation, however, ownership attribution is ambiguous. Users often view generative AI as a tool that facilitates their creative work rather than an independent contributor deserving authorship rights. Recent research supports this pattern: users do not naturally attribute ownership to AI but often assert authorship over AI-generated outputs themselves [30, 41, 60, 61]. This phenomenon is influenced by both users' engagement and system design. Wasi et al. [60] found that while many participants initially deny ownership of unmodified AI-generated content, their sense of ownership increased when they were reminded of their role in generating prompts and editing responses. Similarly, Joshi and Vogel [33] showed that requiring longer prompts helped restore psychological ownership when writing with AI. Importantly, Longoni et al. [41] found that users' belief that AI does not have a true ownership over its content leads to differences in moral judgments regarding plagiarism. Specifically, people apply stricter plagiarism standards to human-created content than AI-generated content. Based on this literature, we hypothesize:

- **H2:** Individuals will attribute greater ownership to the writer when their submitted content closely resembles AI-generated content rather than human-created content.

These differences in perceptions of moral patiency and ownership can affect how individuals evaluate the ethical aspects of using AI-generated content. Longoni et al. [41] demonstrated that participants perceived plagiarism of AI-generated work as less severe than plagiarism of human-created work. This suggests that when users do not perceive AI as either a legitimate owner or a moral patient, their ethical judgments change in a way that reduces their sense of moral responsibility and feelings of guilt. Building on this, we propose that moral patiency and ownership perceptions together mediate moral judgments and feelings of guilt. Specifically, we hypothesize:

- **H3:** The effect of content source (AI vs. human) on feelings of guilt and moral judgments will be mediated by perceptions of moral patiency attributed to the original source and ownership attributed to the writer.

## 2.3 Anthropomorphic Cues and Moral Judgments

As AI becomes increasingly embedded in collaborative work environments, perceptions of AI agency take on heightened significance [32, 56]. Such perceptions are not only grounded in individual psychology but are also shaped by the design of AI systems [34]. These perceptions also unfold within the sociotechnical context of AI systems themselves, whose design features can enable or constrain particular moral judgments and actions, a perspective often described as the moral affordances of artifacts [16].

According to the CASA framework, people attribute social qualities, including agency and responsibility, to machines based on their behavior, especially when AI systems make decisions autonomously [47, 49]. In such contexts, AI may no longer be seen solely as a passive tool but as a more active collaborator, raising expectations about its adherence to moral norms and accountability.

The attribution of agency may be influenced by anthropomorphism, a psychological process where users ascribe human-like qualities to non-human entities [18, 44]. In human-robot interaction, individuals often interact with machines as if they possess minds, attributing intentions, emotions, or even moral qualities to them [1, 15, 59]. Anthropomorphism can be triggered by various design elements, such as names, conversation styles, or facial expressions [37], which can make the AI appear more agentic and socially engaging.

While such cues can enhance users' engagement with AI, prior research also emphasizes that the impact of anthropomorphism depends on the fit between the design cue and the system's actual capabilities. Human-like features that do not correspond to meaningful behavioral or emotional depth may fail to sustain trust or perceived authenticity [5, 31]. Some scholars even caution that anthropomorphic features can operate as dark patterns when they create misleading impressions of agency and moral capacity [46]. These insights suggest the moral implications of anthropomorphism are not universal but contingent on the salience and credibility of the human-like cues employed.

Our study focuses on a minimal-cue context regarding anthropomorphism, testing whether even a simple human-like name can influence perceptions of AI's moral patiency and ownership. In many real-world AI systems, human-like names serve as a common and intentionally deployed anthropomorphic cue. Virtual assistants such as Alexa, Siri, and Cortana, as well as conversational agents that refer to themselves using first-person pronouns (e.g., "I"), routinely incorporate such cues to facilitate social engagement and signal a recognizable identity. Research in HCI and social psychology, including the CASA framework, shows that even minimal human-like cues can elicit social responses without producing full anthropomorphic perception.

When AI systems entirely lack anthropomorphic cues, users may attribute little to no moral patiency or ownership to them. In such cases, users may view AI outputs as unclaimed resources, free to be used or reused without acknowledgement. However, when even minimal anthropomorphic elements are introduced, such as assigning a human-like name, users may begin to recognize traces of agency or authorship, prompting stronger moral or ethical consideration of the AI's contribution. Prior work shows that people attribute moral blame to AI systems when they perceive them as capable of intentions [55], suggesting that subtle cues may trigger moral evaluation processes.

Anthropomorphic cues may reduce moral disengagement by increasing the chance that users perceive AI as having some limited form of "mind." When users hold such perceptions, unethical behavior involving AI, such as plagiarizing AI-generated content, is more likely to be judged as morally wrong. Conversely, when AI is treated as a mere tool, users may rationalize their actions by dismissing the possibility of harm and ownership to AI. Perceiving AI as agentic and capable of meaningful contribution can also lead users to attribute partial ownership to the system, whereas treating it as a passive instrument diminishes both ownership and ethical responsibility, encouraging more permissive use of AI-generated outputs without proper attribution. Based on these insights, we propose the following hypotheses:

- **H4:** The presence of an anthropomorphic cue (i.e., a human-like name for the AI author) will lead to higher perceptions of moral patiency for an AI source compared to an AI source without anthropomorphic cues.
- **H5:** The presence of an anthropomorphic cue (i.e., a human-like name for the AI author) will lead to lower ownership attributions to the writer compared to an AI source without anthropomorphic cues.
- **H6:** Perceptions of moral patiency attributed to the original source and ownership attributed to the writer will mediate the effect of the anthropomorphic cue on feelings of guilt and on moral judgments regarding potential plagiarism.

## 3 Method
### 3.1 Study Design and Procedure

An online experiment was conducted between November and December 2024 with students at a large university in the United States (US) to investigate participants' moral judgments about potential plagiarism involving AI- and human-generated content. Participants were asked to read excerpts from two academic journal manuscripts (title, abstract, and the first paragraph of the introduction) and evaluate their similarity and potential plagiarism. This scenario was modeled on an academic writing context, allowing participants, primarily college students, to engage with a familiar and structured instance of plagiarism where norms about authorship and citation are well understood.

Participants were randomly assigned to one of the three experimental conditions. In the first condition, the original journal article was described as produced using an AI system, specifically a Large Language Model (LLM) like ChatGPT (without any anthropomorphic cues). In the second condition, participants were told that the original journal article was produced using an AI agent named Taylor Lee, powered by LLMs. In the third condition, participants were informed that the original article was created by a human author named Taylor Lee. After receiving the source information, participants reviewed excerpts from the two manuscripts. They were told that the second manuscript was written by a hypothetical author, Sam Davis, which shared approximately 70% of its content with the original manuscript. The similarity between the two manuscripts was pre-tested with a separate sample of participants (n = 18) who rated the content similarity on a scale of 0 to 100%, yielding a mean similarity score of 72.59% ($SD$ = 17.78).

After reviewing the manuscripts, participants completed manipulation and attention check questions, followed by measures of key mediating and outcome variables. Finally, participants provided demographic information, including age, gender, education, and race/ethnicity. Demographic categories reflected conventions commonly used in the US. The study protocol and survey materials were approved by an Institutional Review Board. Participants were recruited through a university department-managed online subject pool, which allows students to volunteer for research studies in exchange for extra credit in eligible courses.

To avoid ambiguity, we use the terms "author" and "writer" separately throughout the manuscript to distinguish the creator of the original text from the person reusing that text. Specifically, we use the term "author" to refer to the entity that created the source manuscript in each condition (a human author, an AI system, or an AI agent with a human-like name). We use the term "writer" to refer to Sam Davis, the individual whose behavior participants evaluated when judging similarity, plagiarism, and ethicality.

## 3.2 Measures

*3.2.1 Perceived Moral Patiency.* Perceived moral patiency was assessed using a scale adapted from previous research by Banks [8], Banks and Bowman [9], and Ladak et al. [40], based on Gray et al.'s [23] conceptual framework. The wording of the items was slightly modified across conditions (indicated in brackets) to reflect the study condition. Participants rated their agreement with four statements assessing the perceived harm caused by copying content, including: "Copying content [generated by an AI system/written by a human] causes significant harm to [the AI system/the AI system, Taylor Lee/Taylor Lee, the person who created it]," "[An AI system/A person] is wronged when its content is plagiarized," "Plagiarizing from [an AI system/a human author] violates its ownership rights," and "[An AI system/a human creator] experiences damage when their content is plagiarized." Responses were recorded on a 7-point Likert scale ranging from 1 (strongly disagree) to 7 (strongly agree), with Cronbach's alpha = .89.

*3.2.2 Perceived Ownership.* Perceived ownership was evaluated using items derived from Longoni et al. [41] and Fyfe [20]. Participants evaluated the degree to which they believed the article submitted by Sam Davis belonged to them and reflected their original work. The two items used were: "To what extent does the article that Sam Davis submitted belong to Sam Davis?" and "To what extent is the article submitted by Sam Davis their own?" Responses were collected using a 7-point Likert scale, ranging from 1 (not at all) to 7 (very much), with a Spearman-Brown coefficient of .85.

*3.2.3 Perceived Feelings of Guilt.* To measure perceived guilt, we used a scale adapted from Giroux et al. [21], drawing on work by Cotte et al. [14]. Participants rated the likelihood that Sam Davis would feel guilty, irresponsible, or accountable for using content created by the AI or human author. The items included: "If I were Sam, I would feel guilty if I published the paper as my own work," "If I were Sam, I would feel irresponsible if I failed to credit [the AI author/the AI author, Taylor Lee/the author, Taylor Lee]," and "If I were Sam, I would feel accountable for using the content created by [the AI author/the AI author, Taylor Lee/the author, Taylor Lee]." Ratings were provided on a 7-point Likert scale from 1 (strongly disagree) to 7 (strongly agree), with Cronbach's alpha = .62.

*3.2.4 Perceived Unethicality.* Perceived unethicality was evaluated using items adapted from Longoni et al. [41]. Participants assessed the extent to which they believed Sam's actions were "unethical," "immoral," or "wrong." Responses were recorded on a 7-point Likert scale, with anchors ranging from 1 (not at all) to 7 (very much), with Cronbach's alpha = .84.

*3.2.5 Perceived Degree of Plagiarism.* Perceived plagiarism was assessed using a questionnaire adapted from Chan et al. [12] and Longoni et al. [41]. Participants rated their agreement with statements regarding whether Sam Davis's actions constituted plagiarism, violated copyright, or represented unlawful or unethical behavior. The items included: "To what extent do you think what Sam Davis did was plagiarism?" "To what extent do you think what Sam Davis did violated the original author's copyright?" "To what extent do you think what Sam Davis did was unlawful?" and "To what extent do you think what Sam Davis did was stealing the original author's work?" Responses were recorded on a 7-point Likert scale ranging from 1 (not at all) to 7 (very much), with Cronbach's alpha = .89.

## 3.3 Manipulation Check

After reviewing the two manuscripts, participants completed manipulation check questions. All participants were asked: "Who authored the Original Article?" to assess the agent manipulation (AI vs. Human). Response options included: "An AI author," "A human author," and "I don't know." Among those assigned to the AI conditions, 91.3% correctly identified the author, while 87.7% of those in the human condition did the same, confirming that the agent manipulation was successful.[1] For participants in the AI conditions, an additional question was asked to assess anthropomorphism manipulation: "Was the AI author described with a specific name?" Response options included: "Yes," "No," and "I don't know." Among those assigned to the AI without an anthropomorphic condition, 84.7% correctly answered, while 92.7% of those in the AI with anthropomorphic condition did the same, suggesting that the manipulation had the intended effect.[2]

## 3.4 Participants

A total of 190 participants completed the online experiment, with 160 participants included in the final analysis after excluding those who failed an attention check question.[3] The final sample sizes[4] for each condition were: AI (n = 59), AI with anthropomorphic cue (n = 55), and human author (n = 46). The mean age of participants was 19.49 years (SD = 1.54, range: 18–24). Regarding gender, 45.6% identified as male, 51.2% as female, 1.3% as non-binary/third gender, 0.6% preferred not to say, and 0.6% chose to self-describe. In terms of ethnicity, 6.9% of participants identified as Hispanic, Latino/a, or of Spanish origin. Participants' racial identities were reported as follows: 65.0% identified as White or Caucasian, 22.5% as Asian, 3.8% as Black or African American, 1.3% as Native American or Alaska Native, 1.3% as Native Hawaiian or Other Pacific Islander.

## 3.5 Data Analysis

We used a one-way multivariate analysis of variance (MANOVA) to examine whether the pattern of dependent variables differed by content source. The correlations of variables used in the analyses are reported in Table 1.

To examine the mediating role of perceived ownership and moral patiency in the relationship between experimental conditions and moral judgments (perceived guilt, unethicality, and plagiarism), we employed Hayes' PROCESS macro (Model 4) [28] to test the indirect effects of experimental conditions on moral judgments through perceived ownership and moral patiency. Figure 1 illustrates the conceptual mediation model.

---

[1] $X^2(2) = 104.49$, $p < .001$
[2] $X^2(2) = 88.94$, $p < .001$
[3] The attention check asked participants, "What was the name of the person who referred to the Original Article and submitted their work to the journal?" The correct answer was "Sam Davis."
[4] A power analysis using GPower 3.1 indicated that for a MANOVA with three experimental groups and five dependent variables, assuming a medium effect size (f2=0.10) alpha level of 0.05, and a power of 0.80, the minimum total sample size required was 87 participants. Our sample size exceeded this minimum to ensure robustness and allow for exclusions.

Table 1: Intercorrelations, means, and standard deviations of measured variables

| Variable | 1 | 2 | 3 | 4 | 5 |
|---|---|---|---|---|---|
| 1. Moral Patiency | – | | | | |
| 2. Ownership | -0.13 | – | | | |
| 3. Guilt | 0.43*** | -0.32*** | – | | |
| 4. Unethicality | 0.48*** | -0.37*** | 0.58*** | – | |
| 5. Plagiarism | 0.52*** | -0.41*** | 0.51*** | 0.74*** | – |
| M | 3.78 | 4.02 | 5.30 | 4.54 | 4.61 |
| SD | 1.72 | 1.49 | 1.23 | 1.46 | 1.34 |

$N = 160$, *** $p < .001$

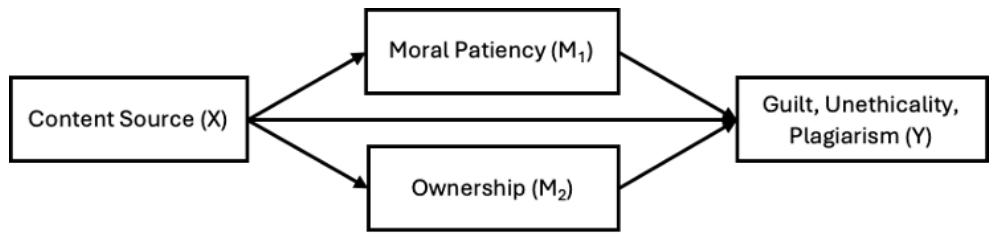

Figure 1: Effects of content source on guilt, unethicality, and plagiarism perceptions through mediators.

Table 2: Observed means and standard deviations by experimental condition

| Variable | Human | AI (No Anthro) | AI (Anthro) | F | p | $\eta p^2$ |
|---|---|---|---|---|---|---|
| Moral Patiency | 5.47 (1.04)[a] | 2.97 (1.47)[b] | 3.24 (1.43)[b] | 51.15 | < .001 | .40 |
| Ownership | 3.58 (1.40)[a] | 4.63 (1.47)[b] | 3.73 (1.41)[a] | 8.72 | < .001 | .10 |
| Guilt | 5.85 (1.09)[a] | 4.93 (1.27)[b] | 5.25 (1.13)[b] | 8.03 | < .001 | .09 |
| Unethicality | 5.26 (1.14)[a] | 4.11 (1.48)[b] | 4.41 (1.49)[b] | 9.27 | < .001 | .11 |
| Plagiarism | 5.18 (0.89)[a] | 4.22 (1.42)[b] | 4.53 (1.42)[b] | 7.29 | < .01 | .09 |

Means with different superscripts (a, b) indicate statistically significant differences based on post hoc comparisons using the Bonferroni adjustment ($p < .05$).

The mediation analyses followed a hierarchical approach. First, the two AI conditions were collapsed into a single category and compared this category against the human condition. This analysis provided an overall assessment of differences in mediators and outcome variables between AI-generated and human-created content. Next, we compared the two AI conditions to evaluate the specific influence of anthropomorphic cues.

## 4 Results

The MANOVA examined the effect of content source (Human, AI with anthropomorphic cue, and AI without anthropomorphic cue) on perceptions related to moral patiency, ownership, guilt, unethicality, and plagiarism. Next, univariate analyses of variance (ANOVAs) were conducted to examine differences in moral patiency (H1, H4) and ownership perceptions (H2, H5) based on content source and the presence of an anthropomorphic cue. Mediation analyses assessed whether moral patiency and ownership mediated the effects of content source on guilt, unethicality, and plagiarism (H3) and whether they mediated the effects of anthropomorphic cues (H6).

The multivariate test revealed a significant multivariate effect of content source on the combined dependent variables, Wilks' $\Lambda = .54$, $F(10, 306) = 11.06$, $p < .001$, $\eta p^2 = .27$, indicating that the type of content source significantly influenced outcome variables. A one-way ANOVA was conducted to examine the effects of content source on each dependent variable. The findings are summarized in Table 2.

### 4.1 The Effect of Content Source

*4.1.1 Effects on Moral Patiency and Perceived Ownership.* Participants perceived significantly greater moral patiency when the original content was created by a human ($M = 5.47$, $SD = 1.04$) compared to AI without an anthropomorphic cue ($M = 2.97$, $SD = 1.47$) and AI with an anthropomorphic cue ($M = 3.24$, $SD = 1.43$), $F(2, 157) = 51.15$, $p < .001$, $\eta p^2 = .39$. In other words, individuals attribute lower moral patiency to an AI author compared to a human author, supporting H1. Pairwise comparisons showed that both AI conditions were rated significantly lower than the human condition ($p < .001$), but there was no significant difference between AI with and without an anthropomorphic cue ($p = .30$). Thus, we did not find evidence supporting that the presence of an anthropomorphic cue (i.e., human-like name) increases moral patiency perception compared to AI without such a cue. Accordingly, H4 was not supported.

There was a significant effect of content source on ownership perceptions, $F(2, 157) = 8.72$, $p < .001$, $\eta p^2 = .10$. Participants attributed significantly higher ownership to the writer when their submitted content closely resembles AI-generated content and when AI was introduced without an anthropomorphic cue ($M = 4.63$, $SD = 1.47$), compared to both the human condition ($M = 3.58$, $SD = 1.40$, $p < .001$) and the AI with an anthropomorphic cue condition ($M = 3.73$, $SD = 1.41$, $p < .01$). When the original content was generated by an AI system without a human-like name, participants perceived a greater ownership by the writer compared to when the writer used human-created content. Thus, H2 was supported. Participants attributed lower ownership to the writer when the AI author had a human-like name compared to when the AI was presented without anthropomorphic cues ($p < .01$), supporting H5. The ownership perceptions associated with the human condition and the AI-with-anthropomorphic-cue condition did not differ significantly ($p = .60$).

Together, these results indicate that participants viewed AI authors as less morally patient than humans and perceived the writer as having greater ownership of text derived from AI systems, particularly when the AI lacked human-like attributes.

*4.1.2 Effects on Guilt, Unethicality, and Plagiarism Perceptions.*
Next, we examined the effects of experimental conditions on the key outcome variables. Establishing these relationships confirms that content source shapes the moral outcomes central to our model, providing a basis for testing the indirect pathways. The effect of content source on feelings of guilt was significant, $F(2, 157) = 8.03$, $p < .001$, $\eta p^2 = .09$. Participants reported greater guilt when dealing with human-created content ($M = 5.85$, $SD = 1.09$) compared to AI-generated content, whether with or without an anthropomorphic cue (AI with an anthropomorphic cue: $M = 5.25$, $SD = 1.13$, $p < .05$; without an anthropomorphic cue: $M = 4.93$, $SD = 1.27$, $p < .001$). No significant difference was observed between the two AI conditions ($p = .15$).

A significant main effect of content source was found for perceived unethicality, $F(2, 157) = 9.27$, $p < .001$, $\eta p^2 = .11$. Participants rated the writer's action as more unethical when dealing with human-created content ($M = 5.26$, $SD = 1.14$), compared to both AI without an anthropomorphic cue ($M = 4.11$, $SD = 1.48$, $p < .001$) and AI with an anthropomorphic cue ($M = 4.41$, $SD = 1.49$, $p < .05$). The AI conditions did not significantly differ ($p = .24$).

Finally, the effect of content source on perceived plagiarism was significant, $F(2, 157) = 7.29$, $p < .01$, $\eta p^2 = .08$. Participants perceived a higher level of plagiarism when the original content was human-created ($M = 5.18$, $SD = 0.89$), compared to AI-generated content without an anthropomorphic cue ($M = 4.22$, $SD = 1.42$, $p < .001$) and AI-generated content with an anthropomorphic cue ($M = 4.53$, $SD = 1.42$, $p < .05$). Again, no significant difference emerged between the two AI conditions ($p = .21$).

Overall, participants judged the writer's actions as less unethical, less guilt-inducing, and less plagiaristic when the source content was AI-generated than when it was human-authored.

### 4.2 Mediation of the Effect of Human vs. AI-Generated Content

To examine the psychological mechanisms driving differences in moral judgments when a human writer uses AI-generated content compared to human-created content, we conducted a series of mediation analyses. Specifically, we tested whether perceived moral patiency and ownership mediate the effect of content source (AI-generated vs. human-created) on moral evaluations, including guilt, perceptions of unethicality, and the degree of plagiarism.

We first examined whether moral patiency and ownership perceptions mediated the effect of content source on feelings of guilt, unethicality, and plagiarism perceptions (H3). In this set of analyses, the independent variable (X) was content source (human vs. AI), with moral patiency ($M_1$) and ownership perceptions ($M_2$) as mediators, and feelings of guilt, perceived unethicality, and perceived plagiarism as the dependent variables (Y). Figure 2a, figure 2b, and figure 2c summarize the results of mediation analyses.

Participants' perceptions of moral patiency and ownership varied significantly depending on whether the second article was derived from AI-generated or human-created content. AI-generated content was perceived as having significantly lower moral patiency than human-created content ($b = -2.37$, $p < .001$), indicating that participants saw AI as incapable of being harmed from its content being copied without giving it credit. Additionally, the writer was perceived as having greater ownership over the final work when they used AI-generated content compared to when they used human-created content ($b = 0.62$, $p < .05$), indicating that AI is not viewed as a legitimate content owner in the same way as a human author. This suggests that participants viewed the writer as having stronger claims to authorship over the final piece when copying from AI compared to copying from another human writer.

When looking at the effect of content source on guilt and moral judgments, our findings indicate that moral concerns about plagiarism were significantly lower when the writer copied from AI-generated content rather than from human-created content. Participants reported lower anticipated guilt for the writer when they used AI-generated content ($b = -0.77$, $p < .001$), perceived their actions as less unethical ($b = -1.01$, $p < .001$) and judged them as involving lower levels of plagiarism ($b = -0.81$, $p < .001$). These results suggest that AI-generated content is treated differently in plagiarism judgments than human-authored content, leading to more lenient moral evaluations.

### 4.3 The Role of Moral Patiency and Ownership Perceptions

Next, we examined whether moral patiency and perceived ownership mediated the effect of content source on guilt, unethicality, and plagiarism perceptions. First, we found that the perceived incapability of AI to be harmed (i.e., lower moral patiency) played a major role in moral judgments. When the writer copied from AI-generated content, participants saw the AI as an entity less capable of being harmed, which contributed to weaker moral evaluations of the plagiarism incident. The indirect effects of content source on moral outcomes through moral patiency were significant for guilt (95% CI [-1.00, -0.23]), unethicality (95% CI[-1.42, -0.55]), and plagiarism perceptions (95% CI[-1.56, -0.77]).

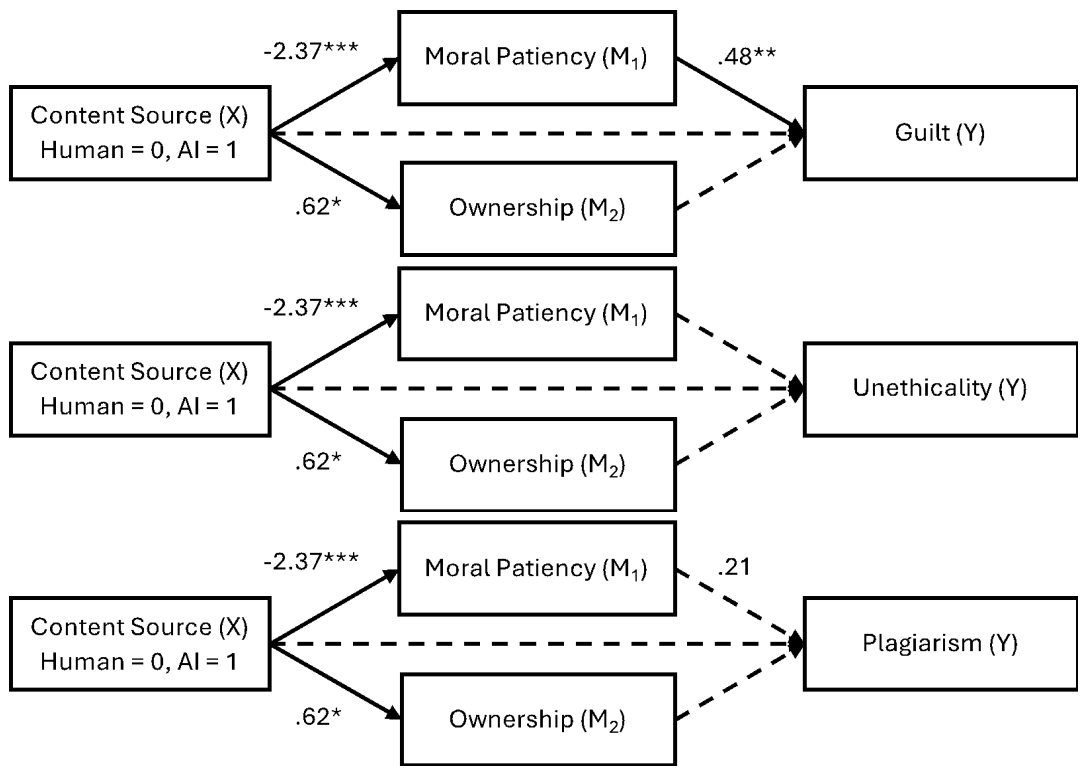

Figure 2: (a) Effects of AI source on feelings of guilt through mediators, $N = 160$, $^{*}p < 0.05$, $^{**}p < 0.01$, $^{***}p < 0.001$, significant and non-significant paths are depicted with solid and dashed lines, respectively. (b): Effects of AI source on perceived unethicality through mediators. (c) Effects of AI source on plagiarism perception through mediators.

The indirect effect of content source on moral outcomes through perceived ownership was statistically significant but smaller in magnitude than the effect through moral patiency. The mediating role of perceived ownership was significant for guilt (95% CI [-0.30, -0.03]), unethicality (95% CI[-0.47, -0.05]), and plagiarism perceptions (95% CI[-0.50, -0.05]).

When both mediators were included in the model, the direct effects of content source on guilt, unethicality, and plagiarism perceptions were no longer significant, suggesting full mediation by moral patiency and perceived ownership. Therefore, H3 was supported. These findings suggest that the moral leniency observed in AI plagiarism judgments reflects differences in how people attribute moral capacity and ownership to human versus AI sources.

### 4.4 Mediation of the Effect of Anthropomorphic Cue

To explore if the same psychological mechanism explains the influence of anthropomorphic cues (H6), we conducted mediation analyses comparing two AI conditions: without an anthropomorphic cue and with an anthropomorphic cue (a named AI author, Taylor Lee). Figure 3a, figure 3b, and figure 3c summarize the results of mediation analyses.

The results showed that adding an anthropomorphic cue influenced perceptions of content ownership but had no significant effect on moral patiency. Specifically, participants perceived the writer as having lower ownership over their final work when the AI was named Taylor Lee compared to when it was not named and presented as an anonymous system ($b = -.90$, $p < .01$). However, moral patiency perceptions did not significantly differ between the two AI conditions ($b = .26$, $p = .34$), suggesting that naming the AI did not change participants' perceptions of whether the AI could be harmed or had moral standing.

We next examined how the presence of an anthropomorphic cue influenced moral judgments. First, the anthropomorphic cue had no significant direct effect on anticipated guilt ($b = -.14$, $p = .86$), indicating that simply naming the AI author did not directly influence whether participants felt the writer would feel guilty. Similarly, the anthropomorphic cue did not significantly affect perceptions of unethicality ($b = -.06$, $p = .95$) nor perceived plagiarism ($b = -.89$, $p = .23$). While giving the AI a human-like name alone did not directly influence moral judgments, mediation analysis revealed statistically significant indirect effects through perceived ownership but not through moral patiency.

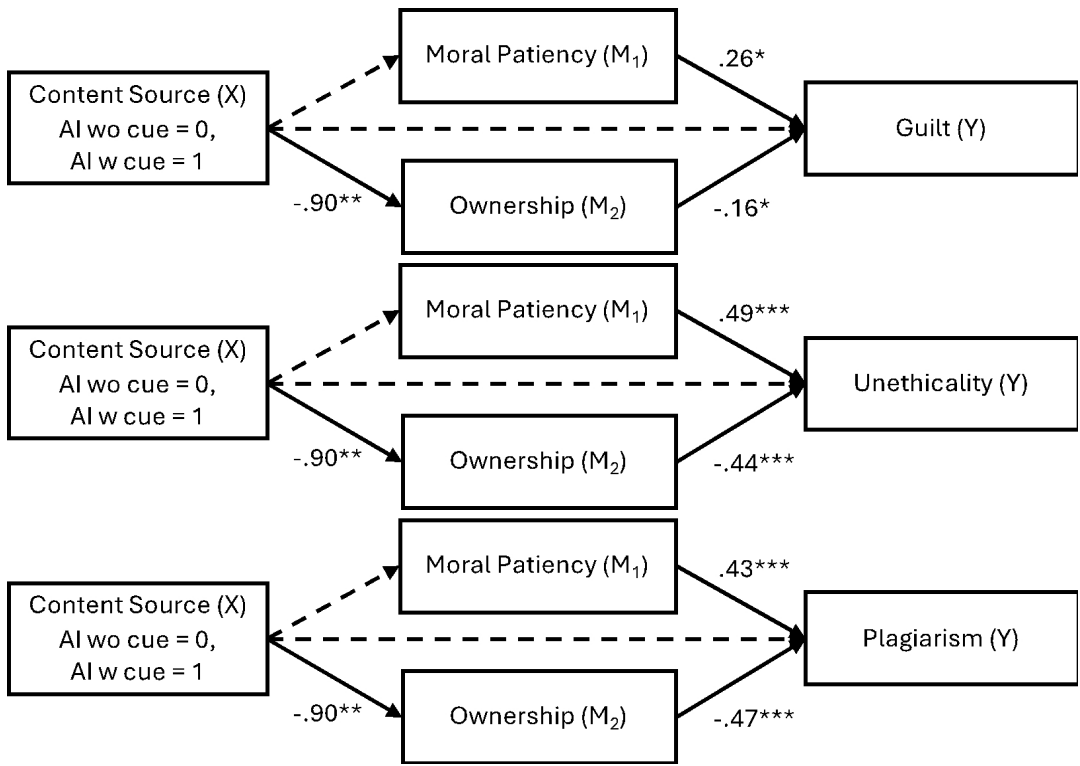

Figure 3: (a) Effects of anthropomorphic cue on feelings of guilt through mediators, $N = 114$, $^*p < 0.05$, $^{**}p < 0.01$, $^{***}p < 0.001$, non-significant paths are depicted with dashed lines. (b) Effects of anthropomorphic cue on perceived unethicality through mediators. (c) Effects of AI source on plagiarism perception through mediators.

Since perceptions of moral patiency did not significantly differ between the two AI conditions, its role as a mediator was not supported. However, perceived ownership significantly mediated the effect of the anthropomorphic cue on guilt, unethicality, and plagiarism perceptions. The indirect effects of the anthropomorphic cue on guilt through ownership (95% CI [0.02, 0.41]), unethicality (95% CI [0.07, 0.71]), and plagiarism perceptions through ownership were significant (95% CI [0.12, 0.74]). This suggests that anthropomorphizing the AI influenced moral judgments not by changing perceptions of AI's ability to experience harm, but by altering ownership perceptions. Thus, H6 was partially supported.

## 5 Discussion

The present research shows that participants judged plagiarism of AI-generated content significantly more leniently than plagiarism of human-authored work. This pattern aligns with growing evidence that AI-generated material occupies an ambiguous moral status, shaping how people justify the use of such content without attribution.

Our findings reveal two psychological mechanisms underlying this leniency. First, participants attributed substantially lower moral patiency to AI than to humans, consistent with Mind Perception Theory [23]. Because AI was not viewed as capable of being harmed, copying from AI was judged to be less morally consequential. Second, participants perceived themselves as having greater ownership over AI-derived content than content taken from a human author, indicating that AI is not viewed as a legitimate author with a rightful claim to intellectual property. Together, these mechanisms reflect a form of moral disengagement [7], in which users minimize perceived harm and amplify their own entitlement to justify ethically questionable behavior.

The minimal anthropomorphic cue used in the study, a human-like name, influenced these judgments in subtle and specific ways. Naming the AI agent did not increase its perceived moral patiency or directly alter guilt, unethicality, or plagiarism perceptions. Instead, the cue indirectly heightened ethical evaluations of the writer's behavior by reducing perceptions of the writer's ownership over the final text. Human-like cues appear to make AI outputs feel more like authored contributions rather than neutral tools, even if they do not fundamentally shift beliefs about AI's moral standing.

At the same time, human-like cues raise their own ethical concerns. Prior work in HCI has shown that human-like cues in AI systems can lead users to over-attribute agency, emotionality, or consciousness to computational systems, sometimes fostering inappropriate trust, dependency, or intimacy [2]. Thus, while minimal anthropomorphic cues may shape ownership attributions in ways that increase accountability, they may also introduce new vulnerabilities and misconceptions. Recognizing both the potential benefits and risks of human-like framing is essential for the responsible design of AI systems. Importantly, these psychological dynamics do not unfold in isolation but within the broader design and sociotechnical affordances of generative AI systems.

Technologies possess moral affordances that can enable, constrain, or subtly guide ethical behavior through their form and functionality [16]. Features of generative AI, such as frictionless content production, obscured data provenance, and minimal cues suggesting sentience or emotional capacity, can make it easier for users to distance themselves from notions of harm, responsibility, or attribution. From this perspective, the moral leniency observed in our study reflects not only judgments about AI's perceived mind but also the affordances of generative tools that shape how users encounter and reuse AI-assisted writing. Recognizing this interplay between user perception and technological affordances highlights the need for ethical frameworks that attend to both user cognition and design characteristics of human-AI co-creative systems.

Together, these findings help clarify the underlying psychological basis of how users judge the ethical status of AI-generated content. By demonstrating that perceptions of moral patiency and ownership systematically influence judgments of guilt, unethicality, and plagiarism, this study provides a foundational account of why disclosure of AI involvement often feels optional or unnecessary to users. These insights enrich ongoing discussions about transparency, accountability, and shared authorship in human–AI co-creation, and offer conceptual grounding for future research on how design interventions, disclosure mechanisms, and normative frameworks might support more responsible engagement with generative AI.

### 5.1 Design Implications

The findings help explain why attribution and disclosure norms remain difficult to establish in AI-assisted writing. Participants' reduced perceptions of AI's moral patiency and heightened sense of ownership over AI-derived text help explain why users may not view acknowledgment of AI involvement as ethically necessary. These cognitive patterns suggest that emerging authorship practices must account for how users intuitively interpret the status of AI-generated contributions within human–AI co-creative workflows.

In many writing tools, AI involvement is embedded in ways that can make the division of labor between human and machine difficult to perceive. When the boundaries between user contributions and AI-generated content are unclear, users may naturally interpret the output as primarily their own. Increasing transparency around how the AI contributes to the writing process, for example, by indicating when content is generated or how prompts shaped the output, may help users form more informed judgments about authorship and attribution in collaborative writing contexts.

Anthropomorphic cues also warrant careful consideration in system design. Although naming the AI shifted ownership perceptions in ways that heightened ethical judgements, human-like framings can introduce unintended effects, including encouraging over-attributions of agency or emotional capacity. Designers and practitioners should therefore carefully balance any benefits of signaling contribution against the risks of implying capabilities that AI systems do not possess.

Overall, these implications highlight opportunities for shaping responsible authorship norms in AI-mediated writing. By understanding how people interpret moral patiency, ownership, and contribution in human–AI collaboration, researchers, educators, and system developers can support clearer expectations and more consistent, ethical attribution practices across diverse writing contexts.

## 6 Limitations and future work

Several limitations of the present research warrant consideration. First, our sample consisted primarily of college students at a US university, which may limit the generalizability of the findings to other populations. This demographic and institutional context may shape moral evaluations of plagiarism. In US higher education, issues surrounding academic integrity and AI use are highly salient, and students' familiarity with related policies and classroom discussions may heighten their sensitivity to misconduct scenarios. Future work should examine whether similar patterns emerge in professional, creative, or cross-cultural contexts where norms of authorship and attribution differ. Second, the study focused on an academic writing context, where issues of authorship and attribution are particularly relevant for participants. While this focus allowed for clear interpretation of moral judgments, plagiarism norms and expectations vary across domains such as journalism, creative writing, and the arts, where conventions of credit, transparency, and creative reuse vary. Future work should investigate whether perceptions of moral patiency and ownership operate similarly in these diverse contexts of AI-mediated creation. Third, broader ethical questions surrounding generative AI extend beyond user plagiarism to include the data practices of AI systems themselves, such as the use of human-created materials for model training without consent. Although this dimension lies outside the empirical scope of the present study, future studies could examine how awareness of these upstream practices influences users' moral evaluations of AI-generated content. Fourth, our work focuses on users' moral evaluations of plagiarism, but this represents only one facet of the ethical challenges posed by generative AI. Other related concerns, such as transparency in disclosing AI involvement, accurate representation of authorship, and the potential for AI-generated content to obscure human contributions, were beyond the scope of the present study. Future research should examine how people reason about these additional ethical dimensions, particularly in settings where authorship ambiguity or lack of transparency may carry meaningful practical consequences. Moreover, while our study examined the effect of a simple anthropomorphic cue (a human-like name), future research could explore whether more elaborate anthropomorphic features, such as conversational style, visual avatars, or emotional expressions, might further influence moral judgments. Additionally, we asked participants to evaluate the moral behavior of a hypothetical third party rather than reflect on their own behavior. Individuals' may judge their own actions differently due to self-serving biases. People may interpret their own ethically questionable behavior more favorably, minimizing responsibility and potential harm. The effects observed in the current study, especially the moral leniency toward AI plagiarism, may be even stronger if participants evaluated their own use of AI-generated content. Future research should examine this possibility by employing self-referential scenarios or behavioral designs in which participants engage in or reflect on their own AI-assisted content creation practices.

## 7 Conclusion

This study sheds light on how individuals navigate the ethical ambiguity surrounding the use of AI-generated content. Our findings reveal that moral leniency toward plagiarism involving AI-generated content is associated with two key psychological mechanisms: diminished perceptions of moral patiency and increased perceptions of content ownership by the human user. While anthropomorphic cues may subtly shift ownership perceptions, they do not alter the underlying belief that AI systems are incapable of experiencing harm. As generative AI tools continue to permeate communication, education, and creative domains, understanding these psychological dynamics is essential for informing ethical guidelines and fostering responsible AI use.

## References

[1] Airenti, G. 2015. The Cognitive Bases of Anthropomorphism: From Relatedness to Empathy. International Journal of Social Robotics. 7, 1 (Feb. 2015), 117–127. https://doi.org/10.1007/s12369-014-0263-x.
[2] Akbulut, C. et al. 2025. All Too Human? Mapping and Mitigating the Risks from Anthropomorphic AI. Proceedings of the 2024 AAAI/ACM Conference on AI, Ethics, and Society (San Jose, California, USA, Feb. 2025), 13–26.
[3] Anthis, J.R. et al. 2025. Perceptions of Sentient AI and Other Digital Minds: Evidence from the AI, Morality, and Sentience (AIMS) Survey. Proceedings of the 2025 CHI Conference on Human Factors in Computing Systems (New York, NY, USA, Apr. 2025), 1–22.
[4] Avey, J.B. et al. 2009. Psychological ownership: theoretical extensions, measurement and relation to work outcomes. Journal of Organizational Behavior. 30, 2 (Feb. 2009), 173–191. https://doi.org/10.1002/job.583.
[5] Aylett, M.P. et al. 2019. The right kind of unnatural: designing a robot voice. Proceedings of the 1st International Conference on Conversational User Interfaces (Dublin Ireland, Aug. 2019), 1–2.
[6] Balle, S.N. 2022. Empathic responses and moral status for social robots: an argument in favor of robot patienthood based on K. E. Løgstrup. AI & SOCIETY. 37, 2 (June 2022), 535–548. https://doi.org/10.1007/s00146-021-01211-2.
[7] Bandura, A. 2011. Moral Disengagement. The Encyclopedia of Peace Psychology. D.J. Christie, ed. Wiley.
[8] Banks, J. 2021. From Warranty Voids to Uprising Advocacy: Human Action and the Perceived Moral Patiency of Social Robots. Frontiers in Robotics and AI. 8, (May 2021). https://doi.org/10.3389/frobt.2021.670503.
[9] Banks, J. and Bowman, N.D. 2023. Perceived Moral Patiency of Social Robots: Explication and Scale Development. International Journal of Social Robotics. 15, 1 (Jan. 2023), 101–113. https://doi.org/10.1007/s12369-022-00950-6.
[10] Brave, S. et al. 2005. Computers that care: investigating the effects of orientation of emotion exhibited by an embodied computer agent. International Journal of Human-Computer Studies. 62, 2 (Feb. 2005), 161–178. https://doi.org/10.1016/j.ijhcs.2004.11.002.
[11] Burgoon, J.K. and Hale, J.L. 1984. The fundamental topoi of relational communication. Communication Monographs. 51, 3 (Sept. 1984), 193–214. https://doi.org/10.1080/03637758409390195.
[12] Chan, C.K.Y. 2023. Is AI Changing the Rules of Academic Misconduct? An In-depth Look at Students' Perceptions of "AI-giarism." arXiv.
[13] Congressional Research Service 2023. Generative artificial intelligence and copyright law. Technical Report #LSB10922. Library of Congress.
[14] Cotte, J. et al. 2005. Enhancing or disrupting guilt: the role of ad credibility and perceived manipulative intent. Journal of Business Research. 58, 3 (Mar. 2005), 361–368. https://doi.org/10.1016/S0148-2963(03)00102-4.
[15] Darling, K. 2021. The new breed: how to think about robots. Allen Lane.
[16] Davis, J.L. 2020. How Artifacts Afford: The Power and Politics of Everyday Things. The MIT Press.
[17] Deterding, S. et al. 2017. Mixed-Initiative Creative Interfaces. Proceedings of the 2017 CHI Conference Extended Abstracts on Human Factors in Computing Systems (Denver Colorado USA, May 2017), 628–635.
[18] Epley, N. et al. 2007. On seeing human: A three-factor theory of anthropomorphism. Psychological Review. 114, 4 (2007), 864–886. https://doi.org/10.1037/0033-295X.114.4.864.
[19] Frazer, R. et al. 2022. Moral Disengagement Cues and Consequences for Victims in Entertainment Narratives: An Experimental Investigation. Media Psychology. 25, 4 (July 2022), 619–637. https://doi.org/10.1080/15213269.2022.2034020.
[20] Fyfe, P. 2023. How to cheat on your final paper: Assigning AI for student writing. AI & SOCIETY. 38, 4 (Aug. 2023), 1395–1405. https://doi.org/10.1007/s00146-022-01397-z.
[21] Giroux, M. et al. 2022. Artificial Intelligence and Declined Guilt: Retailing Morality Comparison Between Human and AI. Journal of Business Ethics. 178, 4 (July 2022), 1027–1041. https://doi.org/10.1007/s10551-022-05056-7.
[22] Gouldner, A.W. 1960. The Norm of Reciprocity: A Preliminary Statement. American Sociological Review. 25, 2 (1960), 161–178. https://doi.org/10.2307/2092623.
[23] Gray, K. et al. 2012. Mind Perception Is the Essence of Morality. Psychological Inquiry. 23, 2 (Apr. 2012), 101–124. https://doi.org/10.1080/1047840X.2012.651387.
[24] Guingrich, R.E. and Graziano, M.S.A. 2024. Ascribing consciousness to artificial intelligence: human-AI interaction and its carry-over effects on human-human interaction. Frontiers in Psychology. 15, (Mar. 2024), 1322781. https://doi.org/10.3389/fpsyg.2024.1322781.
[25] Haidt, J. 2001. The emotional dog and its rational tail: A social intuitionist approach to moral judgment. Psychological Review. 108, 4 (2001), 814–834. https://doi.org/10.1037/0033-295X.108.4.814.
[26] Haidt, J. ed. 2013. The righteous mind: why good people are divided by politics and religion. Vintage Books.
[27] Hancock, J.T. et al. 2020. AI-Mediated Communication: Definition, Research Agenda, and Ethical Considerations. Journal of Computer-Mediated Communication. 25, 1 (Mar. 2020), 89–100. https://doi.org/10.1093/jcmc/zmz022.
[28] Hayes, A.F. 2018. Introduction to mediation, moderation, and conditional process analysis: a regression-based approach. Guilford Press.
[29] Hwang, A.H.-C. et al. 2025. "It was 80% me, 20% AI": Seeking Authenticity in Co-Writing with Large Language Models. Proc. ACM Hum.-Comput. Interact. 9, 2 (May 2025), CSCW122:1-CSCW122:41. https://doi.org/10.1145/3711020.
[30] Jelson, A. and Lee, S.W. 2024. An empirical study to understand how students use ChatGPT for writing essays and how it affects their ownership. arXiv.
[31] Jensen, T. and Khan, M.M.H. 2022. I'm Only Human: The Effects of Trust Dampening by Anthropomorphic Agents. HCI International 2022 – Late Breaking Papers: Interacting with eXtended Reality and Artificial Intelligence (Cham, 2022), 285–306.
[32] Jia, H. et al. 2022. Do We Blame it on the Machine? Task Outcome and Agency Attribution in Human-Technology Collaboration. Proceedings of the 55th Hawaii International Conference on System Sciences (2022).
[33] Joshi, N. and Vogel, D. 2025. Writing with AI Lowers Psychological Ownership, but Longer Prompts Can Help. Proceedings of the 7th ACM Conference on Conversational User Interfaces (Waterloo ON Canada, July 2025), 1–17.
[34] Kadoma, K. et al. 2024. The Role of Inclusion, Control, and Ownership in Workplace AI-Mediated Communication. Proceedings of the 2024 CHI Conference on Human Factors in Computing Systems (New York, NY, USA, May 2024), 1–10.
[35] Khalaf, M.A. 2024. Does attitude towards plagiarism predict aigiarism using ChatGPT? AI and Ethics. (Feb. 2024). https://doi.org/10.1007/s43681-024-00426-5.
[36] Kim, T. et al. 2023. AI increases unethical consumer behavior due to reduced anticipatory guilt. Journal of the Academy of Marketing Science. 51, 4 (July 2023), 785–801. https://doi.org/10.1007/s11747-021-00832-9.
[37] Kim, Y. and Sundar, S.S. 2012. Anthropomorphism of computers: Is it mindful or mindless? Computers in Human Behavior. 28, 1 (Jan. 2012), 241–250. https://doi.org/10.1016/j.chb.2011.09.006.
[38] Kyi, L. et al. 2025. Governance of Generative AI in Creative Work: Consent, Credit, Compensation, and Beyond. Proceedings of the 2025 CHI Conference on Human Factors in Computing Systems (New York, NY, USA, Apr. 2025), 1–16.
[39] Ladak, A. et al. 2025. Robots, Chatbots, Self-Driving Cars: Perceptions of Mind and Morality Across Artificial Intelligences. Proceedings of the 2025 CHI Conference on Human Factors in Computing Systems (New York, NY, USA, Apr. 2025), 1–19.
[40] Ladak, A. et al. 2024. The Moral Psychology of Artificial Intelligence. Current Directions in Psychological Science. 33, 1 (Feb. 2024), 27–34. https://doi.org/10.1177/09637214231205866.
[41] Longoni, C. et al. 2023. Plagiarizing AI-generated Content Is Seen As Less Unethical and More Permissible. Preprint. https://doi.org/10.31234/osf.io/na3wb.
[42] Loomis, J.L. 1959. Communication, the Development of Trust, and Cooperative Behavior. Human Relations. 12, 4 (Nov. 1959), 305–315. https://doi.org/10.1177/001872675901200402.
[43] Lovato, J. et al. 2024. Foregrounding Artist Opinions: A Survey Study on Transparency, Ownership, and Fairness in AI Generative Art. Proceedings of the AAAI/ACM Conference on AI, Ethics, and Society. 7, 1 (Oct. 2024), 905–916. https://doi.org/10.1609/aies.v7i1.31691.
[44] Maeda, T. and Quan-Haase, A. 2024. When Human-AI Interactions Become Parasocial: Agency and Anthropomorphism in Affective Design. Proceedings of

...

the 2024 ACM Conference on Fairness, Accountability, and Transparency (New York, NY, USA, June 2024), 1068–1077.

[45] Malle, B.F. *et al.* 2025. People's judgments of humans and robots in a classic moral dilemma. Cognition. 254, (Jan. 2025), 105958. https://doi.org/10.1016/j.cognition.2024.105958.

[46] Mildner, T. *et al.* 2024. Listening to the Voices: Describing Ethical Caveats of Conversational User Interfaces According to Experts and Frequent Users. Proceedings of the CHI Conference on Human Factors in Computing Systems (Honolulu HI USA, May 2024), 1–18.

[47] Nass, C. and Moon, Y. 2000. Machines and Mindlessness: Social Responses to Computers. Journal of Social Issues. 56, 1 (Jan. 2000), 81–103. https://doi.org/10.1111/0022-4537.00153.

[48] OECD 2025. Intellectual property issues in artificial intelligence trained on scraped data.

[49] Reeves, B. and Nass, C.I. 1996. The media equation: How people treat computers, television, and new media like real people and places. CSLI Publ.

[50] Schecter, A. and Richardson, B. 2025. How the Role of Generative AI Shapes Perceptions of Value in Human-AI Collaborative Work. Proceedings of the 2025 CHI Conference on Human Factors in Computing Systems (New York, NY, USA, Apr. 2025), 1–15.

[51] Schmidt, P. and Loidolt, S. 2023. Interacting with Machines: Can an Artificially Intelligent Agent Be a Partner? Philosophy & Technology. 36, 3 (Sept. 2023), 55. https://doi.org/10.1007/s13347-023-00656-1.

[52] Schroeder, D.A. and Linder, D.E. 1976. Effects of actor's causal role, outcome severity, and knowledge of prior accidents upon attributions of responsibility. Journal of Experimental Social Psychology. 12, 4 (Jan. 1976), 340–356. https://doi.org/10.1016/S0022-1031(76)80003-0.

[53] Shaver, K.G. 1985. The attribution of blame: Causality, responsibility, and blameworthiness. Springer-Verlag.

[54] Sidoti, O. *et al.* 2025. About a quarter of U.S. teens have used ChatGPT for schoolwork – double the share in 2023. Pew Research Center.

[55] Sullivan, Y.W. and Fosso Wamba, S. 2022. Moral Judgments in the Age of Artificial Intelligence. Journal of Business Ethics. 178, 4 (July 2022), 917–943. https://doi.org/10.1007/s10551-022-05053-w.

[56] Sundar, S.S. 2020. Rise of Machine Agency: A Framework for Studying the Psychology of Human–AI Interaction (HAII). Journal of Computer-Mediated Communication. 25, 1 (Mar. 2020), 74–88. https://doi.org/10.1093/jcmc/zmz026.

[57] Swanepoel, D. 2021. Does Artificial Intelligence Have Agency? The Mind-Technology Problem. R.W. Clowes *et al.*, eds. Springer International Publishing. 83–104.

[58] Tanibe, T. *et al.* 2017. We perceive a mind in a robot when we help it. PLOS ONE. 12, 7 (July 2017), e0180952. https://doi.org/10.1371/journal.pone.0180952.

[59] Turkle, S. 2011. Alone together: why we expect more from technology and less from each other. Basic books.

[60] Wasi, A.T. *et al.* 2024. LLMs as Writing Assistants: Exploring Perspectives on Sense of Ownership and Reasoning. arXiv.

[61] Xu, Y. *et al.* 2024. What Makes It Mine? Exploring Psychological Ownership over Human-AI Co-Creations. Nova Scotia. (2024).

[62] Yannakakis, G.N. *et al.* 2014. Mixed-initiative co-creativity. (2014).